\documentclass[journal]{IEEEtran}

\usepackage{cite}
\usepackage{graphicx}
\usepackage{amsmath,amssymb}
\usepackage{array}
\usepackage{url}
\usepackage{stfloats}

\begin{document}

\title{Neural Network Model for OAM Crosstalk due to Turbulence-Induced Tilt and Lateral Displacement}

\author{Mitchell~A.~Cox*, Steven~G.~Makoni, and Ling~Cheng%
\thanks{M. A. Cox, S. G. Makoni and L. Cheng are with the School of Electrical and Information Engineering, University of the Witwatersrand, Johannesburg, South Africa (email: mitchell.cox@wits.ac.za).}%
}

\maketitle

\begin{abstract}
Accurately modelling orbital angular momentum (OAM) mode crosstalk in turbulent environments is challenging yet essential for developing free-space optical systems that employ OAM modes for multiplexing or diversity. Turbulence induces tip/tilt aberrations and lateral displacement, which significantly degrade system performance. Existing analytical models describe the transformation from an input Gaussian mode to an output OAM spectrum; however, our feed forward neural network model generalizes this approach by accounting for the effects of these aberrations on arbitrary input OAM modes. We validate the model experimentally by estimating turbulence-induced tilt and lateral displacement using a dual-camera setup and comparing the estimated spectrum with the actual modal decomposition. With a typical root mean square error of less than 11\%, our results indicate that the model could serve as a reliable source of meta-information for digital signal processing, soft-decision forward error correction or perhaps dynamic mode hopping in future systems.
\end{abstract}


\IEEEpeerreviewmaketitle

\section{Introduction}

Free-space optical (FSO) communication is attracting considerable interest as an alternative or complement to radio-frequency systems and fibre networks due to its high bandwidth potential and cost-effectiveness in scenarios where fibre installation is impractical or prohibitively expensive \cite{imran2024ieee,Trichili2020RoadmapOpticsb,lavery2018}. The rapid growth in data demand has driven the development of advanced multiplexing schemes originally pioneered in fibre communications. Space division multiplexing (SDM) and, in particular, mode-division multiplexing (MDM) offer the capability to transmit multiple independent data channels simultaneously by exploiting orthogonal spatial modes of light \cite{Ciaramella20091.28-Tb/sSystem,Richardson2013Space-divisionFibres,Gibson2004Free-spaceMomentum}.

One promising approach within MDM is the use of beams carrying orbital angular momentum (OAM) or other higher order modes. OAM modes are characterized by a helical phase structure of the form $e^{i\ell\phi}$, where $\ell$ is an unbounded integer known as the OAM charge. This property allows for high-dimensional encoding and a significant increase in spectral efficiency \cite{trichilliSDMreview}.

Atmospheric turbulence poses a significant challenge for OAM-based FSO links by introducing dynamic aberrations---primarily tip and tilt (due to the rotational symmetry of OAM we can consider these as equivalent here) and lateral displacement. These distort the beam's phase front and lead to inter-modal crosstalk \cite{cox2020slt,paterson2005atmospheric,Tyler2009InfluenceMomentum,Anguita2008Turbulence-inducedLink}. While other higher-order mode families such as Hermite--Gaussian (HG), Laguerre--Gaussian (LG) and vector modes have also been investigated for their robustness to turbulence \cite{Peters2023,cox2018diversity,cox2019hglg,Gu2020}, OAM modes remain a convenient test case due to their well-defined azimuthal phase structure and ease of creation and detection. The principles demonstrated here apply broadly and as such we focus on OAM modes due to their relative simplicity, with the understanding that future FSO systems may leverage a wider array of spatial modes.

Turbulence manifests through several degrading effects on optical beams, including beam spreading, scintillation (intensity fluctuations), and beam wandering. In MDM systems specifically, these phenomena translate into crosstalk between spatial modes and mode-dependent loss, which significantly impact system performance. The choice of modes in such systems becomes critical: certain mode sets exhibit differential resilience to turbulence effects, and strategic mode selection can reduce crosstalk or even exploit it constructively through diversity techniques or ``mode averaging'' approaches \cite{drozdov2025}.

An important capability for next-generation FSO systems would be dynamic ``mode hopping'': the ability to adaptively select optimal transmission modes based on current atmospheric conditions. However, implementing such strategies requires accurate, real-time estimation of modal crosstalk under varying turbulence scenarios. This could be done using full modal decompositions with custom optical elements such as mode sorters \cite{berkhout2010efficient}, or even computationally expensive deep learning-based image processing approaches \cite{Cox2022interf}. Here we present an alternative approach. Current analytical models that predict the effect of tip/tilt and lateral displacement on the output OAM spectrum have been developed \cite{Lin:10,Vasnetsov2005AnalysisBeam}; however, these models are limited to Gaussian inputs. For example, Lin \cite{Lin:10} derived a closed-form expression for the OAM mode weight under tip/tilt and lateral displacement:
\begin{align}
c_{\ell}(u_0,v_0) 
&= \exp\Bigl(-\frac{u_0^2 + v_0^2}{4}\Bigr) \notag \\[6pt]
&\quad \times \Biggl( 
\frac{\sqrt{u_0^4 + v_0^4 + 2\,u_0^2\,v_0^2}}
      {u_0^2 + v_0^2}
\Biggr)^{\!\!\ell} \notag \\
&\quad \times I_{\ell}\Biggl(
\frac{\sqrt{u_0^4 + v_0^4 + 2\,u_0^2\,v_0^2}}
     {u_0^2 + v_0^2}
\Biggr),
\label{eq:lin_crosstalk}
\end{align}
where $c_{\ell}$ represents the OAM mode weight, $I_{\ell}(\cdot)$ is the modified Bessel function of the first kind, $u_0 = {2\Delta y}/{\omega_0}$ is the dimensionless displacement parameter, and $v_0 = {(2\pi\,\omega_0\,\sin\alpha)}/{\lambda}$ is the dimensionless tip/tilt parameter. Here, $\Delta y$ is the lateral displacement, $\alpha$ is the tip/tilt angle, $\omega_0$ is the beam waist, and $\lambda$ is the wavelength.

Although this formula effectively captures crosstalk for an initially Gaussian beam, it does not extend to arbitrary input fields (such as pure OAM modes with $\ell\neq 0$) or other higher-order mode superpositions. Similarly, other models in the literature \cite{Klug2021TheFields,Tyler2009InfluenceMomentum} focus on ensemble-averaged behaviour under turbulence and implicitly assume a Gaussian input profile. These limitations underscore the need for a more flexible model that can handle arbitrary input modes and capture instantaneous turbulence effects, particularly to enable adaptive transmission strategies like mode hopping.

Recent advances in data-driven approaches have motivated the use of neural networks to address these challenges. In this work, we introduce a feed-forward neural network (FFNN) model that estimates OAM crosstalk resulting from turbulence-induced tip/tilt and lateral displacement on an instantaneous basis. This approach circumvents the limitations of long-term averaging and analytical models, providing a framework that could enable real-time crosstalk estimation and adaptive mode selection in practical FSO systems. Although our demonstration employs OAM modes for simplicity, the proposed framework can readily extend to other higher-order modes---such as HG and LG---that may be exploited in future SDM/MDM FSO systems.

The remainder of the paper is organized as follows. Section~\ref{sec:background} reviews the fundamentals of the impact of atmospheric turbulence on OAM modes, feed-forward neural networks, and modal decomposition. Section~\ref{sec:nn} details the design, training, and testing of our feed-forward neural network model and the preparation of the associated dataset. Section~\ref{sec:experiments} describes the experimental setup used to validate the model, and Section~\ref{sec:conclusion} concludes with a discussion of potential directions for future research.

\section{Background}
\label{sec:background}
This section provides the fundamental concepts required to understand our approach to modelling turbulence-induced orbital angular momentum (OAM) crosstalk. We review atmospheric turbulence and its effects on optical beams, describe modal decomposition for characterizing spatial modes, and introduce the basics of feed-forward neural networks.

\subsection{Atmospheric Turbulence}

Atmospheric turbulence is a complex physical phenomenon that significantly impacts free-space optical communication systems. It arises from random fluctuations in temperature and pressure within the atmosphere, which in turn create spatial and temporal variations in the refractive index of air. These variations cause beam wandering, scintillation (intensity fluctuations), and wavefront distortions that degrade signal quality.

The statistical properties of these refractive index fluctuations are commonly characterized using the Kolmogorov power spectrum, which describes how turbulent energy is distributed across different spatial scales:
\begin{equation}
\Phi_n(k) = 0.033\, C_n^2\, k^{-11/3}, \quad \frac{1}{L_0} \ll k \ll \frac{1}{l_0},
\end{equation}
where $C_n^2$ is the refractive index structure parameter (a measure of turbulence strength), $k$ is the spatial frequency, $L_0$ is the outer scale of turbulence (typically tens to hundreds of meters), and $l_0$ is the inner scale (typically millimetres). The outer and inner scales represent the upper and lower bounds of the inertial subrange where Kolmogorov's theory accurately describes turbulent behaviour.

For optical beams carrying orbital angular momentum, atmospheric turbulence is particularly problematic because it disrupts the delicate phase structure that defines these modes. Turbulence typically induces low-order aberrations---such as tip, tilt, and defocus---that cause beam misalignments (e.g., lateral displacement) and lead to crosstalk among different OAM modes. These effects are especially pronounced in long-distance free-space links where the cumulative impact of turbulence can completely scramble the modal information \cite{cox2020slt}.

\subsubsection{Noll's Method for Zernike Expansion}
\label{subsec:zernike}

To model the effects of turbulence in a computationally tractable manner, we can decompose the turbulent phase distortion into a set of basis functions. Over a circular aperture, Zernike polynomials provide a convenient orthogonal basis for this purpose:
\begin{equation}
\phi(\rho,\theta) = \sum_{j=1}^{\infty} a_j\, Z_j(\rho,\theta),
\end{equation}
where $\rho$ and $\theta$ are the normalized radial and azimuthal coordinates, and $Z_j(\rho,\theta)$ denotes the $j$th Zernike polynomial. These polynomials are particularly useful because the lower-order terms correspond to familiar optical aberrations such as tilt, defocus, and astigmatism, which are the primary contributors to OAM mode degradation.

Noll's method \cite{Noll1976ZernikeTurbulence} provides a statistical framework for determining the coefficients $a_j$ in turbulent conditions. For Kolmogorov turbulence, the variance of the $j$th coefficient is given by:
\begin{equation}
\langle a_j^2 \rangle = \sigma_j^2 = \kappa\, (n+1) \left(\frac{D}{r_0}\right)^{5/3},
\end{equation}
where $n$ is the radial order of the corresponding Zernike mode, $D$ is the aperture diameter, and $r_0$ is the Fried parameter, which characterizes the spatial coherence of the wavefront. The Fried parameter is defined as:
\begin{equation}
r_0 = \left[ 0.423\, k^2\, C_n^2\, L \right]^{-3/5},
\end{equation}
with $k=\frac{2\pi}{\lambda}$ and $L$ the propagation distance. The constant $\kappa$ is approximately 0.2944 for a circular aperture. 

Turbulence strength can also be approximated with the Strehl Ratio (SR), given below. This is the ratio of the average intensity of a Gaussian beam (i.e. $\ell=0$) with and without turbulence. This is a useful measure that can be used experimentally.
\begin{equation}
\label{eq:SR}
\text{SR} = \frac{\langle I(0) \rangle}{I_0(0)} \approx \frac{1}{\left[1+(D/r_0)^{5/3}\right]^{6/5}},
\end{equation}
where $\langle I(0) \rangle$ is the average intensity at the origin with turbulence, $I_0(0)$ is the intensity at the origin without turbulence, $D$ is the aperture diameter, and $r_0$ is the Fried parameter.

In practical simulations and modelling, phase screens representing atmospheric turbulence are generated by sampling each coefficient $a_j$ from a normal distribution with zero mean and variance $\sigma_j^2$, then summing the weighted Zernike polynomials. This approach allows us to create realistic turbulence scenarios with statistical properties that match those observed in the atmosphere, while maintaining computational efficiency.

\subsection{Modal Decomposition}

Modal decomposition is a fundamental analytical technique in optical communications that allows us to express complex optical fields in terms of simpler, well-understood basis functions. In the context of our work, it provides a powerful framework for analysing how turbulence affects OAM modes \cite{Forbes2020ModalTutorial,peters2025structured}.

An unknown optical field $U(\mathbf{s})$ can be represented as a weighted sum of orthonormal basis functions:
\begin{equation}
U(\mathbf{s}) = \sum_{n=1}^{\infty} c_n\, \Psi_n(\mathbf{s}) = \sum_{n=1}^{\infty} |c_n|\, e^{i\phi_n}\, \Psi_n(\mathbf{s}),
\end{equation}
where the complex coefficients $c_n = |c_n|\, e^{i\phi_n}$ satisfy the normalization condition $\sum_{n}|c_n|^2 = 1$. Each coefficient $c_n$ represents the contribution of the corresponding basis mode $\Psi_n(\mathbf{s})$ to the overall field, with $|c_n|^2$ indicating the fractional power contained in that mode.

These coefficients are computed via the inner product operation, which essentially measures the overlap between the unknown field and each basis function:
\begin{equation}
c_n = \langle \Psi_n | U \rangle = \int \Psi_n^*(\mathbf{s})\, U(\mathbf{s})\, d\mathbf{s}.
\end{equation}

In experimental settings, this mathematical operation is physically realized through a technique known as match filtering. A spatial light modulator (SLM) displays a hologram that, when illuminated by the field under test, transforms the desired mode component into a Gaussian beam that can be coupled into a single-mode fibre and measured. By sequentially displaying holograms corresponding to different basis modes, we can extract the complete modal spectrum of the field.

This process is particularly valuable for studying turbulence effects on OAM modes because it reveals how power is redistributed among modes due to atmospheric distortions. In an ideal, turbulence-free scenario, an input OAM mode would maintain its modal purity, resulting in a spectrum with a single non-zero coefficient. However, when turbulence is present, the spectrum broadens as power leaks into adjacent modes, a phenomenon known as crosstalk. By quantifying this crosstalk through modal decomposition, we can assess the severity of turbulence effects and develop strategies to mitigate them.

\subsection{Feed-Forward Neural Networks}

Feed-forward neural networks are one of the early machine learning models that can approximate complex, non-linear relationships between inputs and outputs. Unlike traditional analytical approaches with fixed mathematical formulations, neural networks learn patterns directly from data, making them particularly well-suited for modelling phenomena like turbulence-induced OAM crosstalk where closed-form solutions are limited and simulations are computationally intensive.

The fundamental building block of an FFNN is the artificial neuron, which performs a weighted sum of its inputs followed by a (usually) non-linear activation function. By organizing these neurons into successive layers, the network can progressively transform input features into increasingly abstract representations, ultimately producing the desired output prediction.

For a network with $L$ layers, the forward computation can be expressed mathematically as:
\begin{align}
\hat{\mathbf{y}} 
&= f^{(L)}\Bigl( W^{(L)}\, f^{(L-1)}\Bigl( \cdots f^{(1)}\Bigl( W^{(1)}\,\mathbf{x} + \mathbf{b}^{(1)} \Bigr) \cdots \nonumber\\[5pt]
&\quad + \mathbf{b}^{(L-1)} \Bigr) + \mathbf{b}^{(L)} \Bigr).
\end{align}
In this expression, $\mathbf{x}$ represents the vector of input features, such as turbulence parameters including tilt angles and lateral displacement in our context. For each layer $l$, the terms $W^{(l)}$ and $\mathbf{b}^{(l)}$ denote the weights and biases, respectively. The activation function $f^{(l)}(\cdot)$, often chosen from ReLU, sigmoid, or tanh, is used to introduce non-linearity into the computation. Together, these components transform input features into the final output $\hat{\mathbf{y}}$, effectively capturing complex data patterns.

The network's ability to model complex relationships stems from its training process, during which the weights and biases are iteratively adjusted to minimize a loss function. For regression tasks such as estimating modal crosstalk coefficients, common loss functions include the mean squared error (MSE):
\begin{equation}
\text{MSE} = \frac{1}{N}\sum_{i=1}^{N}\left(y_i - \hat{y}_i\right)^2,
\end{equation}
or the root mean squared error (RMSE) with $N$ samples:
\begin{equation}
\text{RMSE} = \sqrt{\frac{1}{N}\sum_{i=1}^{N}\left(y_i - \hat{y}_i\right)^2}.
\end{equation}

In this work, we use MSE as the loss function during training because it is differentiable and penalizes larger errors more heavily, which aids in effective weight optimization via back propagation. For model evaluation and reporting purposes, RMSE is often preferred since it is expressed in the same units as the target variables, making it more interpretable.

A crucial aspect of our network design is the output layer's activation function. Since we are predicting modal power distributions (which must be non-negative and sum to one), we employ a softmax activation function:
\begin{equation}
\label{eq:softmax}
\text{softmax}(x_i) = \frac{e^{x_i}}{\sum_{j} e^{x_j}}.
\end{equation}
The softmax function converts raw network outputs into a probability distribution, ensuring that our predicted modal powers satisfy the physical constraints inherent to power-normalized decompositions. This choice of activation function embeds domain knowledge directly into the network architecture, improving both the physical plausibility and accuracy of our predictions.

\section{Neural Network Model}
\label{sec:nn}

\begin{figure*}[tb]
\centering
\includegraphics[width=1\linewidth]{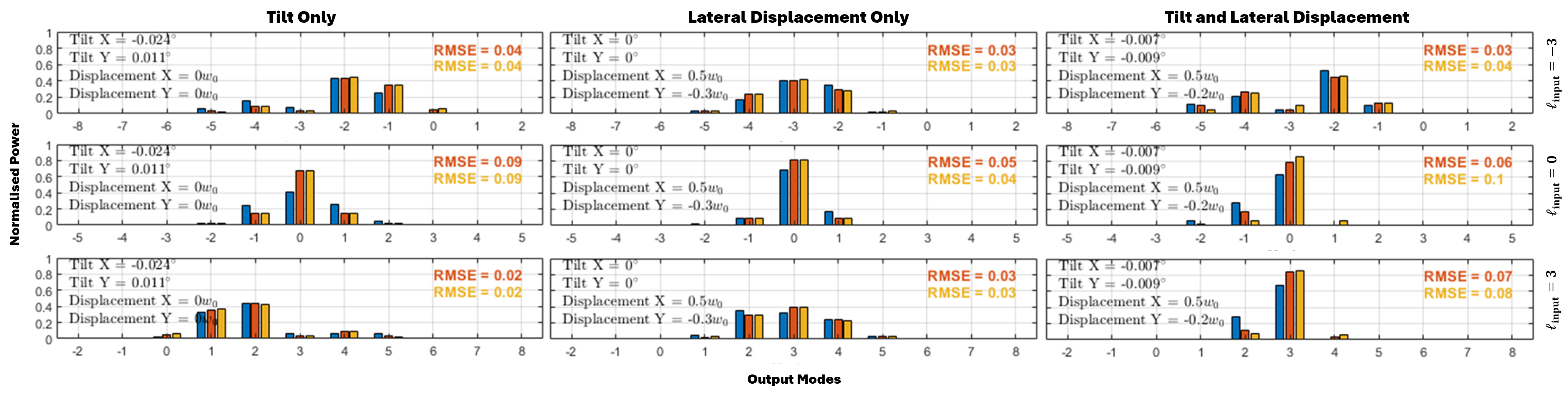}
\caption{Comparison of experimentally measured (blue), simulated (orange), and neural network-estimated (yellow) OAM spectra for various misalignment conditions with the input mode $\ell_{\mathrm{input}}$ being the centre x-axis label. Each subplot corresponds to a different combination of tip/tilt and lateral displacement values, with RMSE values indicated for each case. The model demonstrates high accuracy across different misalignment scenarios, effectively capturing the expected crosstalk behaviour.}
\label{fig:mlcheck}
\end{figure*}

In this section, we present our feed-forward neural network model for estimating OAM crosstalk due to turbulence-induced tip/tilt and lateral displacement. The model maps the input OAM mode and measured misalignment parameters to a predicted OAM spectrum, enabling real-time implementation.

Once trained, the FFNN performs a single forward pass to infer the OAM spectrum based on measured tip/tilt, displacement, and the known input OAM mode. By including various misalignment scenarios in the dataset, the model generalizes well to different conditions, accommodating both isolated and combined misalignments.

Since the network's output is constrained by a softmax function, the predicted OAM spectrum adheres to physical constraints, preventing negative values and ensuring that power is properly distributed across modes (i.e. energy conservation).

\subsection{Data Preparation and Features}

The dataset is generated with simulations due to the large quantity of training data required, and verified using experiments with similar parameters. Each of the twenty thousand samples consists of a transmitted OAM mode, $\ell_{\text{input}}$, which takes discrete values in the range $[-5,5]$. To preserve its categorical nature and avoid implying an ordinal relationship, it is one-hot encoded. The dataset also includes tip/tilt and lateral displacement measurements along the x and y directions. Tip/tilt values range from approximately $-0.25^\circ$ to $0.25^\circ$, while displacement values span $[-1.5\omega_0,\,1.5\omega_0]$, where $\omega_0$ is the beam waist. This range was chosen to capture expected crosstalk under typical misalignment conditions. For stronger turbulence, where displacement and tilt are more pronounced, a larger range would be necessary. In summary, each input sample is represented by a 15-dimensional feature vector, consisting of four normalized continuous features and an 11-dimensional one-hot encoded vector. 

To ensure robustness across different misalignment types, the dataset is divided into three subsets: (i) displacement-only, (ii) tip/tilt-only, and (iii) combined displacement and tip/tilt. To prevent scale disparities from affecting the training process, z-score normalization is applied to the continuous variables (tip/tilt and displacement). 

\subsection{Output Representation}

The network outputs an 11-dimensional vector, normalized with a softmax (see Eq.~\ref{eq:softmax}), representing the power distribution of the OAM spectrum. This spectrum is centred on the input mode, meaning that for a transmitted mode $\ell_{\text{input}}$, the network predicts power values for the adjacent modes spanning $[\ell_{\text{input}}-5,\ell_{\text{input}}+5]$. This formulation ensures that the model captures the expected crosstalk distribution while maintaining consistency with physical modal decomposition results.

\subsection{Network Architecture and Training}

The FFNN (implemented using Python and PyTorch) consists of an input layer with 15 nodes, two hidden layers with 32 neurons each, and an output layer with 11 neurons. The hidden layers use the ReLU activation function to introduce non-linearity, while the output layer remains linear. A softmax function is applied to the raw output to enforce physical constraints, ensuring that all power values are non-negative and sum to one.

An initial manual search was conducted to explore different learning rates, loss functions (MSE, RMSE, MAE), and optimizers (SGD, Adam, RMSProp) to determine their impact on convergence. A subsequent grid search within the most promising parameter range led to the final model configuration, which follows.

The network was trained using MSE as the loss function and optimized with Adam at a learning rate of 0.001. Training is performed with a batch size of 32 over 250 epochs, using a 70:30 train-validation split. These hyper-parameters consistently provided the lowest validation loss while avoiding over fitting.

\subsection{Validation} 

To evaluate the performance of the FFNN model, we validated its predictions against both simulated and experimental datasets. Fig.~\ref{fig:mlcheck} presents a comparison of the experimentally measured, simulated, and neural network-estimated OAM spectra for different misalignment conditions and input modes, $\ell_\mathrm{input} = -3$, 0, and 3, as illustrative examples.

The model performs well across individual misalignment conditions (tilt-only, displacement-only) and combined cases, accurately predicting the expected crosstalk behaviour. The average RMSE values for the validation set (30\% of the data) are 0.008, 0.012, and 0.021 for tilt-only, displacement-only, and combined cases, respectively. The experimentally measured spectra closely match both the simulated and estimated spectra, with minor discrepancies attributed to small experimental misalignments.

These results confirm that the model is robust and suitable for real-time OAM crosstalk estimation in systems where measurements of tilt and lateral displacement are available. In practice, these can be obtained using standard low-cost sensors such as cameras or quadrant photodiodes in FSO systems. 

\section{Crosstalk Estimation in Turbulence}
\label{sec:experiments}

An open question remains: in the presence of the higher-order effects of atmospheric turbulence, will simple estimations of tilt and lateral displacement, when fed through this model, still yield an accurate OAM spectrum estimation? Having validated our neural network model under controlled conditions, we next experimentally examine its performance in turbulence, without retraining or fine-tuning.

\subsection{Experimental Setup}

\begin{figure}[tb]
\centering
\includegraphics[width=0.7\linewidth]{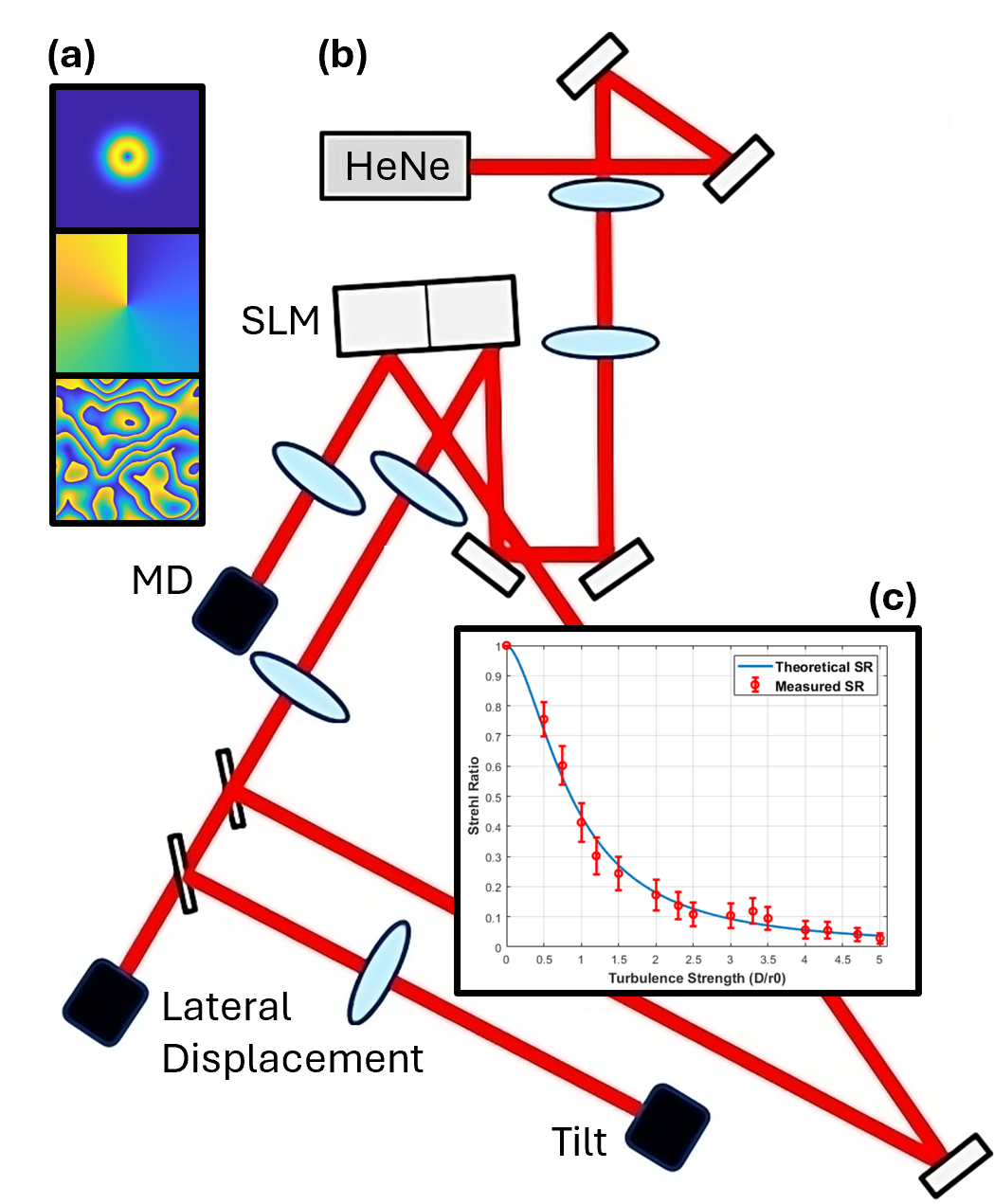}
\caption{Experimental setup (b) for turbulence-induced crosstalk estimation showing an example OAM intensity, phase and turbulence phase screen in (a). Turbulence is simulated by adding Kolmogorov phase screens to the transmitted mode, with calibration shown in the inset (c). Tip/tilt and lateral displacement are estimated using two cameras, and modal decomposition is performed using a second spatial light modulator.}
\label{fig:turbSetup}
\end{figure}

The experiment employed a dual-SLM configuration as illustrated in Fig.~\ref{fig:turbSetup}. The first spatial light modulator (SLM) encoded both a known OAM mode and a Kolmogorov phase screen generated with the method described in Sec.~\ref{subsec:zernike} using 250 Zernike terms, effectively simulating atmospheric turbulence with controlled statistical properties. To ensure the accuracy of our simulated turbulence, we verified the phase screens using measured Strehl ratios which closely match the theoretical prediction shown as an inset in Fig.~\ref{fig:turbSetup}. The beam was then measured with a dual-camera arrangement to estimate tilt and lateral displacement from their centroids. Following these measurements, modal decomposition was performed using a second SLM to determine the actual OAM spectrum resulting from the turbulence-induced distortions.

\subsection{Measurement Procedure}

Our experimental protocol involved generating one hundred random turbulence realizations for each transmitted OAM mode within the range $\ell_{\text{input}} \in [-5,5]$. These realizations spanned various turbulence strengths, with $D/r_0$ values ranging from 0 to 3, specifically $D/r_0=[0, 0.5, 0.7, 1, 1.1, 1.5, 2, 2.3, 2.7, 3]$. For each realization, we performed the following steps:
\begin{enumerate}
\item Encoded the input OAM mode and turbulence phase screen on the first SLM.
\item Approximated tilt angle and lateral displacement values from the two camera images.
\item Used these parameters as inputs to our previously trained neural network model.
\item Performed modal decomposition using the second SLM to determine the actual OAM spectrum.
\item Compared the neural network prediction with the experimentally measured spectrum.
\end{enumerate}

Crucially, we used the neural network model without any retraining, testing its ability to generalize from the controlled conditions of its training data to the more complex scenarios presented by simulated atmospheric turbulence.

\subsection{Results and Analysis}

\begin{figure}[tb]
\centering
\includegraphics[width=1\linewidth]{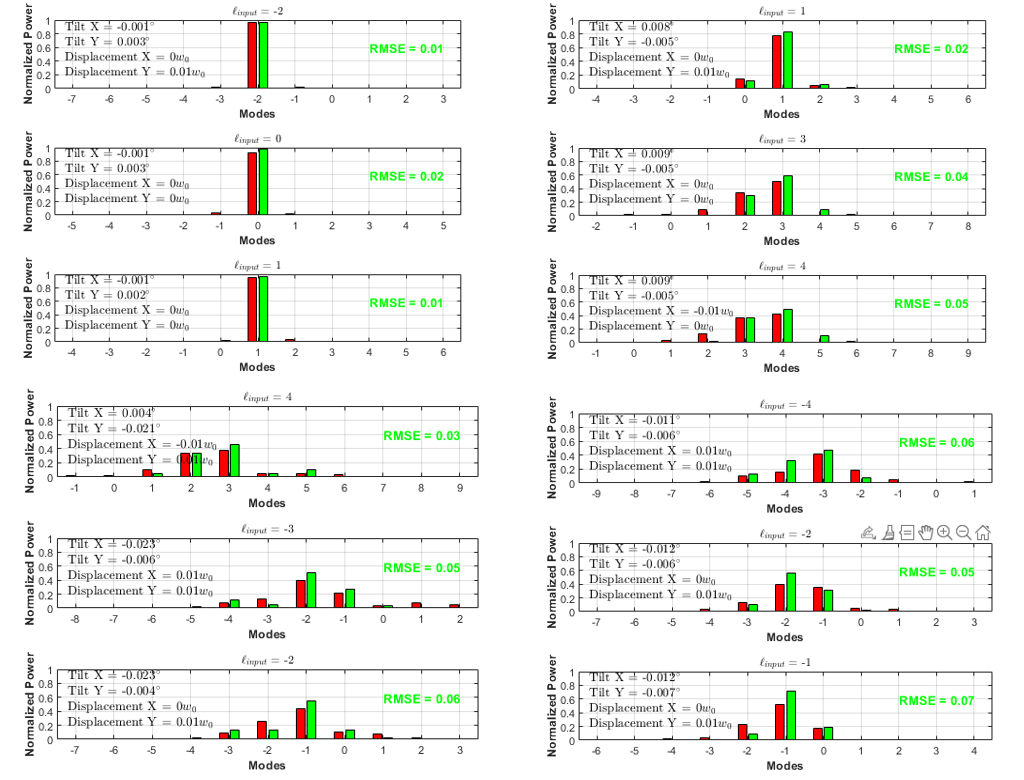}
\caption{Comparison between neural network-estimated (green) and experimentally measured (red) OAM spectra under various turbulence strengths for representative transmitted modes. RMSE values quantify estimation accuracy for each scenario.}
\label{fig:turbresults}
\end{figure}

Fig.~\ref{fig:turbresults} presents representative examples comparing the experimentally measured and neural network-estimated OAM spectra under various turbulence conditions. The results demonstrate that despite the presence of modelled higher-order turbulence effects, the neural network maintained strong predictive performance. Root mean square errors ranged from approximately 0.01 for weak turbulence scenarios to 0.11 for stronger turbulence conditions, with a mean of 0.07 across all 11000 measurements, indicating the model's ability to capture the dominant effects of turbulence on modal crosstalk.

As expected, the estimation accuracy decreased somewhat with increasing turbulence strength, reflecting the growing contribution of higher-order aberrations not explicitly accounted for in our model. However, even under strong turbulence conditions, the model provided useful approximations of the OAM spectrum, maintaining RMSE values below 0.11. With this method of turbulence generation and arrangement, it is difficult to simulate significant lateral displacement, and so in these results the crosstalk is mostly caused by tilt aberrations.

Nevertheless, these results confirm the feasibility of estimating OAM mode crosstalk under realistic turbulence conditions using straightforward measurements of tip/tilt and lateral displacement. The approach provides a computationally efficient and practically implementable technique for real-time crosstalk estimation in turbulence-affected MDM systems, potentially enabling adaptive transmission strategies that could significantly enhance system performance.

\section{Conclusion}
\label{sec:conclusion}

In this work, we introduce an accurate feed-forward neural network model for estimating OAM mode crosstalk resulting from tip/tilt and lateral displacement. Our approach addresses the limitations of computationally complex simulations or existing analytical models by generalizing to arbitrary input OAM modes and providing instantaneous estimations suitable for real-time applications.

In addition, we experimentally demonstrate that our model can accurately predict OAM crosstalk spectra under various turbulence conditions. Inputs to the model are simple measurements of tilt and lateral displacement. Despite the presence of higher-order unmodelled aberrations, the model achieves root mean square errors of less than 11\%, and on average 7\%, for a range of turbulence strengths up to $D/r_0=3$.

These results highlight the potential of our approach to provide computationally efficient and reliable meta-information for digital signal processing, soft-decision forward error correction in real-time OAM-based MDM systems or even potential utility for adaptive transmission strategies such as mode hopping, where the system could dynamically select optimal transmission modes based on current atmospheric conditions. Such capabilities could significantly improve the resilience and throughput of next-generation FSO communication systems.

Future work could explore extending the neural network model to include additional measurable parameters to further improve accuracy in environments with stronger turbulence, or perhaps incorporating predictive capabilities using a recurrent, LSTM-, or transformer-based neural network \cite{Briantcev2023BeamNetworks}. Additionally, investigating the application of this approach to other higher-order mode families---such as Hermite-Gaussian or Laguerre-Gaussian modes---could yield insights into optimizing mode selection for turbulence resilience in next-generation FSO systems.



\bibliographystyle{IEEEtran}
\bibliography{refs}

\begin{IEEEbiographynophoto}{Mitchell Arij Cox}
(Senior Member, IEEE) received his PhD in Electrical Engineering in 2020, focusing on Structured Light in Turbulence. He co-founded the Wits OC Lab, where he applies his expertise in structured light and his proudly-embraced "nerdy engineer" skills to pioneer solutions for long-range, low-cost free-space optical communications. These solutions often involve the creative use of low-cost components—often in ways not originally intended by their manufacturers—and the application of machine learning, evolving into optical neuromorphic computing, to advance low-cost FSO technologies. Dr. Cox was honoured with the 2024 Meiring Naudé Medal from the Royal Society of South Africa, the 2023 Friedel Sellschop Award, and holds a position as a 2023 Optica Ambassador. With his wide-ranging and deep interests, Dr. Cox is committed to using his research to help bridge the digital divide. He aims to make high-speed internet access available globally, beginning with his home in South Africa—tipping his hat to the hard working taxpayers who keep the lasers on.
\end{IEEEbiographynophoto}

\begin{IEEEbiographynophoto}{Steven Gamuchirai Makoni}
received the B.Sc. degree in Electrical and Information Engineering from the University of the Witwatersrand, Johannesburg, South Africa, in 2021, and the M.Sc. degree (with distinction) in Electrical Engineering from the same institution in 2024. His research focused on orbital angular momentum (OAM) modes in turbulent free-space optical (FSO) systems. His research interests include mode-division multiplexing (MDM) in free-space communications, wireless communications, software development, and neural network applications in artificial intelligence. He is also passionate about teaching and student mentorship.
\end{IEEEbiographynophoto}

\begin{IEEEbiographynophoto}{Ling Cheng}
(Senior Member, IEEE) received the B.Eng. degree (cum laude) in electronics and information from Huazhong University of Science and Technology (HUST) in 1995, and the M.-Ing. degree(cum laude) in electrical and electronics and the D.-Ing. degree electrical and electronics from the University of Johannesburg (UJ), in 2005 and 2011, respectively. In 2010, he joined the University of the Witwatersrand, where he was promoted to a Full Professor in 2019. He has been a Visiting Professor with five universities and the principal advisor for over 40 full research post-graduate students (11PhDs). He has published more than 150 research papers in journals and conference proceedings. His research interests are in telecommunications and artificial intelligence. He received the Chancellor’s Medals in 2005 and 2019, the National Research Foundation ratings in 2014 and 2020, and the IEEE ISPLC 2015 Best Student Paper Award was made to his Ph.D. student in Austin. He is the Vice-Chair of IEEE South African Information Theory Chapter. He serves as an associate editor for three journals.
\end{IEEEbiographynophoto}

\end{document}